# Low-Cost Carriers in Aviation: Significance and Developments

Bruno Felipe de Oliveira
Alessandro V. M. Oliveira✈
Aeronautics Institute of Technology, São José dos Campos, Brazil
✈ Corresponding author. E-mail: alessandro@ita.br.

*Abstract*: **This paper aims to discuss the impacts of low-cost airlines on the air transport market and, in particular, to present the most recent findings from the specialized literature in this field. To this end, several papers published on the topic since 2015 were selected and analyzed. Based on this analysis, the main subjects addressed in the studies were categorized into five groups: (i) impacts of low-cost airlines on competing carriers; (ii) impacts on airports; (iii) general effects on air transport demand; (iv) effects on passengers' choice processes; and (v) broader effects on geographical regions.**

## I. Introduction

Since the deregulation of civil aviation in the United States in 1978, the growth of low-cost carriers (LCCs) has attracted the attention of various stakeholders in the sector. This is largely due to the significant impact these airlines have had on the expansion of the aviation industry. In 2015, LCCs transported more than 980 million passengers, representing 28% of all passengers on scheduled flights worldwide.

This result stems from the "low-cost, low-fare" business model, in which airlines offer limited services to passengers through an unbundling strategy — separating the essential air transport service from additional services such as baggage handling, in-flight meals, entertainment systems, or advance seat selection. By adopting this approach and/or increasing operational efficiency through high aircraft utilization and a standardized fleet, airlines can reduce operating costs, which may translate into lower ticket prices.

Applying this business model, Southwest Airlines — one of the largest low-cost airlines in the world and a global benchmark in this segment — has been responsible for billions of dollars in annual savings for consumers on domestic fares. In practice, this translates into lower average ticket prices and a subsequent increase in air travel demand (Beckenstein and Campbell, 2017). Through such strategies, airlines are able to attract price-sensitive passengers, thereby expanding the air transport sector.

Following the Civil Aviation Deregulation Act in the U.S., academic research began to focus on competition in the airline market, particularly on pricing and airline entry into routes. Numerous studies have been conducted in these areas since deregulation, and the number of investigations has grown as other regions around the world have implemented their own civil aviation deregulation policies.

One line of research examines the effects of market entry and competitors' responses to LCC entry. The literature consistently shows that a low-cost carrier effectively reduces average fares upon entering the market — or even when it merely threatens to enter a route (Windle and Dresner, 1999; Morrison, 2001; Goolsbee and Syverson, 2008; Brueckner et al., 2013).

In this context, the primary objective of this study is to shed light on and assess recent research concerning the effects of LCCs not only on airfares but also on other participants in the civil aviation industry. To this end, this study reviews recent publications on the topic, discusses their main conclusions, and identifies potential avenues for future investigation.

## II. Low-Cost Carriers

The expansion of the airline industry is closely tied to the growth of the low-cost business model. In general, a low-cost airline has the following characteristics: it serves short-haul routes; uses regional or secondary airports; operates point-to-point routes; offers limited or no customer loyalty programs; provides limited passenger services; conducts a high proportion of ticket sales online; maximizes aircraft utilization; and maintains a standardized fleet. When an airline operates according to some or all of these characteristics, it reduces operating costs, allowing it to adopt a low-fare strategy. By lowering fares, a low-cost carrier (LCC) can attract price-sensitive passengers who might otherwise use other modes of transportation or choose not to travel at all. The resulting increase in flight demand due to reduced airfares is known in the industry as the "Southwest Effect."

This term was coined by Bennet and Craun (1993) of the U.S. Department of Transportation to describe the increase in air travel that followed Southwest Airlines' entry into new markets after the deregulation of the U.S. civil aviation industry. According to them, the rise in air travel demand occurs in three stages. First, Southwest's entry increases the supply of air travel in a market while offering lower fares than its competitors. Second, competing airlines reduce their own fares to remain competitive and avoid losing passengers to Southwest. Third, due to downward price pressure caused by the increased flight supply and competition with a low-cost carrier, air travel demand grows in markets where Southwest operates.

Since then, several studies have examined the Southwest Effect over time. Dresner et al. (1996) analyzed the impact of Southwest's entry on airlines operating on adjacent routes or at airports near those where Southwest began service. The authors found that even in these cases of indirect competition, prices fell and passenger traffic increased significantly, expanding the understanding of the Southwest Effect. A more recent example is Beckenstein and Campbell (2017), who showed that Southwest was responsible for an average reduction of $45 in ticket prices on routes it entered in 2016. This amount represented a 15% decrease compared to the prices charged by competitors, and as a result of this reduction in average airfares, the number of air trips on those routes increased by 30%.

In addition to examining the Southwest Effect on air travel demand, several studies have also focused on the competitive responses of airlines to Southwest's entry. Ito and Lee (2003), for example, analyzed how airlines operating under the hub-and-spoke model responded to Southwest's entry on routes connected to their hubs. They found that although some airlines reacted with sharp price reductions and aggressive capacity expansion, the average response of incumbents was to accommodate the new entrant by aligning their prices with Southwest's rather than engaging in a price war. Daraban and Fournier (2008) likewise observed that incumbents reduced fares after Southwest's entry and further contributed to the literature by providing evidence that incumbents also cut prices in anticipation of an LCC's entry, and that fares remained lower even after Southwest exited the market. They also confirmed that prices on a given route decrease when Southwest operates on an adjacent route.

However, Southwest is not the only LCC in the world. Founded as Air Southwest in 1967 and renamed Southwest Airlines in 1971, the company has become the largest low-cost airline globally, serving as a benchmark for many other LCCs in different markets. Since then, several studies have analyzed and compared the effects of new low-cost airlines on the market with those attributed to the Southwest Effect.

Windle and Dresner (1999) examined the impact of ValuJet's entry into Delta's hub and analyzed price changes on routes not entered by the LCC. They found that Delta reduced fares on competitive routes connected to its hub in response to ValuJet's competition but found no evidence of fare increases on non-competitive routes to offset revenue losses, suggesting that governments should encourage LCC entry to promote consumer welfare. Brueckner et al. (2013) investigated airline competition and the impacts of low-cost entry on airfares. Their results showed that competition with an LCC significantly affects fares, whether occurring between airport pairs, at adjacent airports, or even through potential competition, confirming and extending previous findings in the literature.

In recent years, with new market strategies and business models, some airlines have exhibited a stronger Southwest Effect than Southwest Airlines itself. Wittman and Swelbar (2013) found that certain low-cost carriers had an even greater impact on reducing average fares than Southwest and suggested that experts should adopt the term "JetBlue Effect," referring to one of the LCCs analyzed in their study.

Within this context of LCC presence in the civil aviation market, this paper discusses recent research in the literature and examines how the most current findings contrast with the classic conclusions regarding the effects of LCCs on different participants in the air transport market, divided into: (i) competing airlines; (ii) airports; (iii) general air transport demand; (iv) passenger choice behavior; and (v) geographic region.

## III. IMPACT OF LCCs ON COMPETING AIRLINES

There are several ways to reshape the landscape of any commercial market, and one of them is through the entry of a new competitor. New entrants can affect existing companies by capturing part of their market share, thereby reducing expected profits. In an effort to defend their position, incumbents often respond by lowering prices or increasing operational efficiency to minimize losses (Porter, 2008).

This general concept from the business world applies well to the aviation industry. As previously discussed, reducing airfares is one way airlines respond to competition from Southwest Airlines or other low-cost carriers. Although this topic has been widely studied since the deregulation of civil aviation in the United States, new research continues to examine it from different perspectives.

Some of the most recent studies on this subject are by Asahi and Murakami (2017) and Ren (2020), both of whom revisited the impact of an LCC on ticket prices by focusing on Southwest Airlines itself. The rationale is that, according to some scholars, Southwest is no longer a strictly low-cost carrier—given its dominant position in the U.S. airline market—but rather a hybrid between a low-cost and a traditional airline. Thus, it is worth investigating whether it still exerts the same influence on competitors' prices as in the past. The authors conclude that Southwest continues to lower average ticket prices on the routes it enters, though the reductions are less pronounced than before. For example, on routes dominated by incumbents or involving direct flights between major national hubs, Southwest's price impact tends to be weaker.

In addition to studies focusing on Southwest, other research has examined LCCs in different markets. Chen (2017) and Zhang et al. (2018), for instance, analyzed the cases of China and Australia, respectively. Both found that the presence of LCCs on routes leads to fare reductions, consistent with classical literature. However, they also noted that in some cases the decrease was less significant, as many of the routes studied showed only a limited presence of LCCs, reducing their potential impact on ticket prices.

Bachwich and Wittman (2017) explored the effects of ultra low-cost carriers (ULCCs)—airlines that minimize operating costs even further than traditional low-cost carriers such as Southwest, allowing them to offer substantially lower fares. They found that both LCCs and ULCCs can reduce average ticket prices on a route, but ULCCs have a stronger effect. However, they also concluded that due to their more aggressive strategies, ULCCs tend to exit markets more quickly than LCCs, often before demand for this type of service fully matures.

Still on the topic of ticket prices, Zou et al. (2017) examined the indirect effect of LCCs on airfares by analyzing baggage check-in fees. They found that the presence of an ULCC or LCC on a route or at an adjacent airport forces competitors that charge for checked baggage to lower their ticket prices, once again corroborating the classic literature on the effects of LCCs on fares.

In addition to price impacts, as highlighted by Porter (2008), the presence of a competitor can compel existing companies to modify their operations or even their business strategies—a hypothesis that has also been tested in recent aviation research. Sun (2015) found that the presence of LCCs on a route increases the differentiation of aircraft departure times as airlines seek to avoid direct competition or peak hours while offering a more differentiated product to passengers. Along similar lines, Mohammadian et al. (2019) reported that competing airlines also adjust the type of aircraft used and the frequency of flights operated on a route. Bendinelli et al. (2016) presented results suggesting that the presence of an LCC may lead incumbent airlines to internalize delay costs.

Regarding the impacts on the operations and strategies of incumbent airlines, Pearson et al. (2015) conducted an exploratory study in the Asian market to assess the strategic capacity of traditional airlines to compete with the growing number of LCCs in the region. They concluded that Northeast Asia is the area where airlines are least prepared to face competition from low-cost carriers, and therefore network carriers in this region should strengthen their strategic capacity both to improve responsiveness to LCCs and to enhance overall performance.

Finally, recent studies have examined the impact of LCCs on airlines operating charter flights. Wu (2016) and Castillo-Manzano et al. (2017) reached similar conclusions, consistent with previous literature: the presence of low-cost carriers has led to a contraction of the charter airline market, forcing these companies to modify their business models—by selling seats on scheduled flights—or to transition into low-cost carriers themselves.

## IV. IMPACT OF LCCs ON AIRPORTS

To develop any aviation market, three elements are essential: the airport, the airline, and the demand for the service. Airports and airlines can be said to have a symbiotic relationship, and those that maximize the potential of their partners often become benchmarks for competitors—whether airports or airlines. In this context, the following section presents research analyzing the impact of an LCC's presence at an airport.

Several studies have attempted to relate the presence of an LCC to potential increases in an airport's operational and/or financial efficiency. Some, such as Yokomi et al. (2017) and Zuidberg (2017), found no evidence that LCC presence leads to greater efficiency. Yokomi et al. (2017), for example, concluded that LCC flights generate lower marginal profits compared to those of traditional airlines, while Zuidberg (2017) argued that the impact of an LCC on airport profitability is almost negligible.

However, other studies have identified positive effects associated with low-cost carriers, such as improvements in financial efficiency (Button et al., 2017), operational efficiency (Martini et al., 2020), or both (Augustyniak et al., 2015). Thus, it is evident that the literature lacks consensus regarding the impact of LCCs on airport efficiency.

One possible explanation for this divergence may lie in the selection of airports analyzed. Tavalaei and Santalo (2019) examined the competitive strategies and efficiencies of airports and found that "pure" airports—those serving exclusively low-cost or traditional airlines—are more efficient than "hybrid" airports, which serve both types.

Beyond exploring the effects of LCCs on airport efficiency, several studies have analyzed the relationship between low-cost airlines and airport connectivity. In this context, connectivity is defined as an indicator of network concentration—that is, the ability to move passengers from one point to another with as few connections as possible, without fare increases, and with minimal connection times and maximum convenience, resulting in benefits for passengers.

Zhang et al. (2017) found that the presence of an LCC has a positive effect on airport connectivity, particularly for tourist or regional airports. This connectivity can even serve as a source of airport revenue by offering facilitation services to passengers making self-connections (Chang et al., 2019)—a market segment still relatively unexplored but with considerable potential for future growth (Cattaneo et al., 2017).

Finally, Jimenez et al. (2017) examined the broader impacts of LCCs on the airport system and concluded that the growth of low-cost airlines is an important driver of physical airport infrastructure development, being closely linked to the expansion or construction of new airports in a given market.

## V. IMPACT OF LCCs ON AIR TRANSPORT DEMAND

In addition to influencing airports, low-cost airlines also have a strong impact on air travel demand, as demonstrated in previous studies on the Southwest Effect. Recent research revisiting this phenomenon has analyzed the current demand-generating potential of LCCs and generally confirmed that their presence increases passenger volumes at airports (Rolim et al., 2016; Boonekamp et al., 2018). However, in some markets, this potential remains limited due to the insufficient number of low-cost flights available (Valdes, 2015; Tsui and Fung, 2016).

Beyond studies examining the effect of LCC presence on overall flight demand at airports or within regions, recent research has also focused on a more specific type of demand—tourism. Alsumairi and Tsui (2017) found that the availability of low-cost flights was one of the main drivers behind the increase in tourist numbers and the overall development of air transport in Saudi Arabia. Similarly, Álvarez-Díaz et al. (2019) reported that in regions characterized by a large diaspora and strong tourist outflow, the presence of low-cost airlines continues to attract visitors, making these carriers a key factor in regional economic development.

## VI. IMPACT OF LCCs ON PASSENGER DECISION-MAKING

Previously, we discussed the role and effects of low-cost carriers (LCCs) on air travel demand. Here, demand is analyzed from a more granular perspective, focusing on how the presence of an LCC can influence passengers' choices of airline, airport, or even travel destination.

Paliska et al. (2016) examined how airports offering low-cost flights can attract passengers. Although LCCs typically appeal to leisure travelers due to their higher price sensitivity, the study found that at Trieste Airport in Italy, low-cost flights attract a greater proportion of business travelers. The main reason for this preference is that LCCs offer point-to-point flights to major European hubs, eliminating the need for connections. Another interesting finding from Paliska et al. (2016) is that passengers living in border regions tend to use airports located in their home countries, thereby reducing the effect of LCC presence in these areas.

In addition to influencing airport choice, LCCs can also affect passengers' choice of airline. Yang (2016), for example, analyzed the case of Northeast Asia and found that passengers in this region show a stronger preference for foreign low-cost carriers, indicating that better-known LCCs have a greater impact on demand than local ones. Saffarzadeh et al. (2016) studied the Iranian market and concluded that, although passengers there prefer low-cost flights, these airlines still need to provide a certain level of service and comfort. In line with this, Kim (2015) investigated perceived value differences between low-cost and traditional flights, finding that passengers are more critical when evaluating the quality of service provided by LCCs. Even when these airlines offer good service, there is no guarantee of passenger loyalty in future bookings. Hunt and Truong (2019) analyzed the specific case of low-cost, long-haul (LCLH) airlines, showing that these carriers are transforming the transatlantic travel market by offering affordable fares and gradually gaining market share, despite ongoing issues with comfort and convenience.

Recent studies have also explored how LCC presence affects different passenger profiles. Cattaneo et al. (2016), for instance, examined the case of students in Italy to determine whether the presence of an airport offering low-cost flights could influence their choice of university. The results confirmed that proximity to an airport with low-cost service is indeed an important factor in students' decisions. Clavé et al. (2015) studied tourists' destination choices and found that while the presence of a low-cost airline can influence destination selection, not all tourists make their decisions based solely on it. Borhan et al. (2017) analyzed how low-cost flights can encourage Libyan car drivers to shift to air transport, demonstrating that LCC availability can be a key motivator for modal change.

Finally, Valdes and Gillen (2018) investigated how LCCs can impact passenger welfare. According to classical economic theory, the mere presence of an LCC should enhance social welfare by lowering airfares. Their study found that in several projected scenarios, the introduction of LCC flights at an airport did indeed increase population welfare. However, in other cases, this effect did not materialize due to market abuse by dominant carriers. Therefore, to ensure that LCC entry genuinely promotes social welfare, a detailed market analysis is necessary rather than simply encouraging these airlines to enter routes indiscriminately.

## VII. IMPACT OF LCCs ON A GEOGRAPHICAL REGION

In addition to examining the effects of LCCs on passenger decisions, some studies have explored how low-cost airlines can influence certain regions beyond the tourism sector. Bowen Jr. (2016), for instance, investigated how the presence of a low-cost carrier in Southeast Asia can enhance the accessibility of peripheral regions to the global air transport market, concluding that LCCs play a crucial role in the growth of secondary airports and the broader regions they serve. Following a similar line of reasoning, Taumoepeau et al. (2017) analyzed the air transport market in Oceania and found that, given the region's unique characteristics—such as small economies, lower population density, and geographically isolated communities—the presence of a hybrid LCC could be key to fostering aviation development in the area.

## VIII. FINAL CONSIDERATIONS

And after all, why should we care about low-cost airlines? As shown in this paper, in a dynamic sector such as air transport, the entry of a new airline can significantly disrupt the existing market landscape. However, when this new entrant operates under a low-cost business model, the effects of its entry can be amplified and felt across different segments of the air transport industry. Classic studies on LCC impacts have primarily examined how these airlines influence airfares and regional flight demand. Following the success of Southwest Airlines in the U.S. market, the low-cost model has been replicated worldwide, undergoing adaptations as new carriers introduce their own interpretations of the concept. Within this highly dynamic context, this study compiled and discussed recent findings in the literature concerning the effects of LCCs on various stakeholders in the air transport market—namely, competitors, airports, general demand, passengers, and geographic regions.

Regarding the effects of LCCs on rival airlines, research has shown that low-cost carriers compel competitors to react in order to maintain their market positions, either by reducing airfares or by adapting operations to improve efficiency. For charter airlines, recent studies have reaffirmed previous findings indicating a substitution effect, in which LCCs gradually replace charter services.

This study also revealed that the effects of LCCs on airports remain inconclusive, suggesting that future research could focus on their impact on airport revenues and connectivity. Additionally, it was found that LCCs continue to stimulate overall demand for air travel, demonstrating that the Southwest Effect remains valid today—both for the general public and for tourist routes.

Finally, several studies reviewed here indicated that the presence of LCCs is a significant factor in passenger decision-making, influencing choices within the air transport sector, such as airlines and airports, as well as beyond it, including universities and tourist destinations. Moreover, LCCs have been shown to contribute to the development of regions previously underserved by air transport.

## ACKNOWLEDGEMENTS

The first author gratefully acknowledges the Coordination for the Improvement of Higher Education Personnel (CAPES), grant code 0001. The second author thanks the São Paulo Research Foundation (FAPESP), grant no. 2020/06851-6 and 2024/016160, and the National Council for Scientific and Technological Development (CNPq), grant no. 305439/2021-9. He also extends his gratitude to ITA colleagues Mauro Caetano, Marcelo Guterres, Evandro Silva, Giovanna Ronzani, Rogéria Arantes, Cláudio Jorge Pinto Alves, Mayara Murça, and Paulo Ivo Queiroz. Any remaining errors are the authors' responsibility.

## REFERENCES

Alsumairi, M., & Tsui, K. H. (2017). A case study: The impact of low-cost carriers on inbound tourism of Saudi Arabia. J. Air Transport Management, 62, 129-145.
Álvarez-Díaz, M., González-Gómez, M., & Otero-Giráldez, M. S. (2019). Low cost airlines and international tourism demand. The case of Porto's airport in the northwest of the Iberian Peninsula. Journal of Air Transport Management, 79, 101689.
Asahi, R., & Murakami, H. (2017). Effects of Southwest Airlines' entry and airport dominance. Journal of Air Transport Management, 64, 86-90.
Augustyniak, W., López-Torres, L., & Kalinowski, S. (2015). Performance of Polish regional airports after accessing the European Union: Does liberalisation impact on airports' efficiency?. Journal of Air Transport Management, 43, 11-19.
Bachwich, A. R., & Wittman, M. D. (2017). The emergence and effects of the ultra-low cost carrier (ULCC) business model in the US airline industry. Journal of Air Transport Management, 62, 155-164.
Beckenstein, A. R., & Campbell, B. (2017). Public benefits and private success: the southwest effect revisited. Darden Business School Working Paper, n. 206.
Bendinelli, W. E., Bettini, H. F., & Oliveira, A. V. (2016). Airline delays, congestion internalization and non-price spillover effects of low cost carrier entry. Transportation Research Part A: Policy and Practice, 85, 39-52.
Bennett, R. D., & Craun, J. M. (1993). The airline deregulation evolution continues: The Southwest effect. Office of Aviation Analysis, U.S. Dep. of Transportation.
Boonekamp, T., Zuidberg, J., & Burghouwt, G. (2018). Determinants of air travel demand: The role of low-cost carriers, ethnic links and aviation-dependent employment. Transportation Research Part A: Policy and Practice, 112, 18-28.
Borhan, M. N., Ibrahim, A. N. H., Miskeen, M. A. A., Rahmat, R. A. O., & Alhodairi, A. M. (2017). Predicting car drivers' intention to use low cost airlines for intercity travel in Libya. Journal of Air Transport Management, 65, 88-98.
Bowen Jr, J. T. (2016). "Now everyone can fly"? Scheduled airline services to secondary cities in Southeast Asia. J. Air Transport Management, 53, 94-104.
Brueckner, J., Lee, D., & Singer, E. (2013). Airline competition and domestic US airfares: A comprehensive reappraisal. Economics of Transportation, 2(1), 1-17.
Button, K., Kramberger, T., Grobin, K., & Rosi, B. (2018). A note on the effects of the number of low-cost airlines on small tourist airports' efficiencies. Journal of Air Transport Management, 72, 92-97.
Castillo-Manzano, J. I., Castro-Nuño, M., López-Valpuesta, L., & Pedregal, D. J. (2017). Measuring the LCC effect on charter airlines in the Spanish airport system. Journal of Air Transport Management, 65, 110-117.

Cattaneo, M., Malighetti, P., Paleari, S., & Redondi, R. (2017). Evolution of the European network and implications for self-connection. Journal of Air Transport Management, 65, 18-28.

Cattaneo, M., Malighetti, P., Paleari, S., & Redondi, R. (2016). The role of the air transport service in interregional long-distance students' mobility in Italy. Transportation Research Part A: Policy and Practice, 93, 66-82.

Chang, Y., Lee, W., & Wu, C. (2019). Potential opportunities for Asian airports to develop self-connecting passenger markets. J. Air Transp. Manag., 77, 7-16.

Chen, R. (2017). Competitive responses of an established airline to the entry of a low-cost carrier into its hub airports. J. Air Transport Management, 64, 113-120.

Clavé, S. A., Saladié, Ò., Cortés-Jiménez, I., Young, A. F., & Young, R. (2015). How different are tourists who decide to travel to a mature destination because of the existence of a low-cost carrier route?. Journal of Air Transport Management, 42, 213-218.

Daft, J., & Albers, S. (2015). An empirical analysis of airline business model convergence. Journal of Air Transport Management, 46, 3-11.

Daraban, B., & Fournier, G. M. (2008). Incumbent responses to low-cost airline entry and exit: A spatial autoregressive panel data analysis. Research in Transportation Economics, 24(1), 15-24.

Dresner, M., Lin, J. S. C., & Windle, R. (1996). The impact of low-cost carriers on airport and route competition. J. Transport Economics and Policy, 309-328.

Dobruszkes, F., Givoni, M., & Vowles, T. (2017). Hello major airports, goodbye regional airports? Recent changes in European and US low-cost airline airport choice. Journal of Air Transport Management, 59, 50-62.

Goolsbee, A., & Syverson, C. (2008). How do incumbents respond to the threat of entry? Evidence from the major airlines. The Quarterly journal of economics, 123(4), 1611-1633.

Hunt, J., & Truong, D. (2019). Low-fare flights across the Atlantic: Impact of low-cost, long-haul trans-Atlantic flights on passenger choice of Carrier. Journal of Air Transport Management, 75, 170-184.

Ito, H., & Lee, D. (2003). Incumbent responses to lower cost entry: evidence from the US airline industry. Brown University Department of Economics Paper, (2003-22).

Jarach, D., Zerbini, F., & Miniero, G. (2009). When legacy carriers converge with low-cost carriers: Exploring the fusion of European airline business models through a case-based analysis. Journal of Air Transport Management, 15(6), 287-293.

Jimenez, E., Claro, J., de Sousa, J. P., & de Neufville, R. (2017). Dynamic evolution of European airport systems in the context of Low-Cost Carriers growth. Journal of Air Transport Management, 64, 68-76.

Kim, Y. (2015). Assessing the effects of perceived value (utilitarian and hedonic) in LCCs and FSCs: Evidence from South Korea. Journal of Air Transport Management, 49, 17-22.

Martini, G., Scotti, D., Viola, D., & Vittadini, G. (2020). Persistent and temporary inefficiency in airport cost function: An application to Italy. Transportation Research Part A: Policy and Practice, 132, 999-1019.

Morrison, S. A. (2001). Actual, adjacent, and potential competition estimating the full effect of Southwest Airlines. Journal of Transport Economics and Policy (JTEP), 35(2), 239-256.

Oliveira, B. F., & Oliveira, A. V. M. (2022). An empirical analysis of the determinants of network construction for Azul Airlines. Journal of Air Transport Management, 101, 102207.

Osterwalder, A., & Pigneur, Y. (2010). Business model generation: a handbook for visionaries, game changers, and challengers. John Wiley & Sons.

Paliska, D., Drobne, S., Borruso, G., Gardina, M., & Fabjan, D. (2016). Passengers' airport choice and airports' catchment area analysis in cross-border Upper Adriatic multi-airport region. Journal of Air Transport Management, 57, 143-154.

Pearson, J., O'Connell, J. F., Pitfield, D., & Ryley, T. (2015). The strategic capability of Asian network airlines to compete with low-cost carriers. Journal of Air Transport Management, 47, 1-10.

Porter, M. E. (2008). The five competitive forces that shape strategy. Harvard business review, 86(1), 25-40.

Ren, J. (2020). Fare impacts of Southwest Airlines: A comparison of nonstop and connecting flights. Journal of Air Transport Management, 84, 101771.

Rolim, P. S., Bettini, H. F., & Oliveira, A. V. (2016). Estimating the impact of airport privatization on airline demand: A regression-based event study. Journal of Air Transport Management, 54, 31-41.

Saffarzadeh, M., Mazaheri, A., Tari, M. Z., & Seyedabrishami, S. (2016). Analysis of Iranian passengers' behavior in choosing type of carrier in international air travel to East Asia. Journal of Air Transport Management, 56, 123-130.

Sun, J. Y. (2015). Clustered airline flight scheduling: Evidence from airline deregulation in Korea. Journal of Air Transport Management, 42, 85-94.

Taumoepeau, S., Towner, N., & Losekoot, E. (2017). Low-Cost Carriers in Oceania, Pacific: Challenges and opportunities. Journal of Air Transport Management, 65, 40-42.

Tavalaei, M. M., & Santalo, J. (2019). Pure versus hybrid competitive strategies in the airport industry. Transportation Research Part A: Policy and Practice, 124, 444-455.

Tsui, W. H. K., & Fung, M. K. Y. (2016). Analysing passenger network changes: The case of Hong Kong. Journal of Air Transport Management, 50, 1-11.

Urban, M., Klemm, M., Ploetner, K. O., & Hornung, M. (2018). Airline categorisation by applying the business model canvas and clustering algorithms. Journal of Air Transport Management, 71, 175-192.

Valdes, V. (2015). Determinants of air travel demand in Middle Income Countries. Journal of Air Transport Management, 42, 75-84.

Valdes, V., & Gillen, D. (2018). The consumer welfare effects of slot concentration and reallocation: A study of Mexico City International Airport. Transportation Research Part A: Policy and Practice, 114, 256-269.

Windle, R., & Dresner, M. (1999). Competitive responses to low cost carrier entry. Transp. Research Part E: Logistics and Transportation Review, 35(1), 59-75.

Wittman, M. D., & Swelbar, W. S. (2013). Evolving trends of US domestic airfares: The impacts of competition, consolidation, and low-cost carriers.

Wu, C. (2016). How aviation deregulation promotes international tourism in Northeast Asia: A case of the charter market in Japan. Journal of Air Transport Management, 57(C), 260-271.

Yang, C. W. (2016). Entry effect of low-cost carriers on airport-pairs demand model using market concentration approach. Journal of Air Transport Management, 57, 291-297.

Yokomi, M., Wheat, P., & Mizutani, J. (2017). The impact of low cost carriers on non-aeronautical revenues in airport: An empirical study of UK airports. Journal of Air Transport Management, 64, 77-85.

Zhang, Y., Sampaio, B., Fu, X., & Huang, Z. (2018). Pricing dynamics between airline groups with dual-brand services: The case of the Australian domestic market. Transportation Research Part A: Policy and Practice, 112, 46-59.

Zhang, Y., Zhang, A., Zhu, Z., & Wang, K. (2017). Connectivity at Chinese airports: The evolution and drivers. Transportation Research Part A: Policy and Practice, 103, 490-508.

Zuidberg, J. (2017). Exploring the determinants for airport profitability: Traffic characteristics, low-cost carriers, seasonality and cost efficiency. Transportation Research Part A: Policy and Practice, 101, 61-72.